\documentclass[11pt]{article}
\usepackage{moriond,epsfig}

\usepackage{amsmath,amssymb}

\begin{document}
\begin{flushright}
 {\footnotesize  FR-PHENO-2012-009, ITP-UU-12/17, SFB/CPP-12-22, SPIN-12/15, 
 TTK-12-18, TUM-HEP-838/12}
\\
\hfill\\
May 4, 2012
 \end{flushright}
\vspace*{1.5cm}

\title{NNLL threshold resummation for the total top-pair production cross section}

\author{Martin Beneke$^{a,b}$, Pietro Falgari$^c$, Sebastian Klein$^b$, 
  Jan Piclum$^{a,b}$, Christian Schwinn$^d$}

\address{\hfill\\
a: Physik Department T31, Technische Universit\"at M\"unchen, James-Franck-Stra\ss e, D - 85748 Garching, Germany\\[.2cm]
b: Institut f\"ur Theoretische Teilchenphysik und 
Kosmologie,
RWTH Aachen University,\\ D--52056 Aachen, Germany\\[.2cm]
c:  Institute for Theoretical Physics and Spinoza Institute,
Utrecht University,\\ 3508 TD Utrecht, The Netherlands\\[.2cm]
d: Albert-Ludwigs Universit\"at Freiburg, 
Physikalisches Institut, 
D-79104 Freiburg, Germany
}

\maketitle\abstracts{
We present predictions for the total top-quark pair production cross
section at the Tevatron and the  LHC with $7,8$ and $14$ TeV centre-of-mass energy, including the resummation of  threshold
logarithms and Coulomb corrections through next-to-next-to-leading
logarithmic order, and $t\bar t$ bound-state contributions. The remaining
theoretical and PDF uncertainties and prospects for the measurement of the
top mass from the total cross section are discussed.
\\[.2cm]
 {\scriptsize\it Based on a talk given by CS at the 47th
 Rencontres de Moriond on QCD and High Energy Interactions, March
 10-17, 2012, La Thuile, Italy.}
}

\section{Introduction}
After the discovery of the top quark at the Tevatron and the initial determination of its properties like mass, decay
width, and the coupling to other particles, the LHC is
currently opening the door to precision studies with hundreds of
thousands of top-antitop pairs produced per year. One of the key
observables is the top-quark pair production cross-section that has now
been measured with an accuracy of seven percent both at Tevatron and
LHC.
 At the LHC, the combinations of measurements in different channels yield the results~\cite{exp-talks}
\begin{equation}
 \sigma_{t\bar t} 
= 
\begin{cases}
177\phantom{A} \pm 3\phantom{A}\text{(stat.)} \pm 8\phantom{AB} \text{(syst.)} \pm 7\phantom{A}\text{ (lumi.) pb},&\text{ATLAS}\\
165.8\pm 2.2 \text{ (stat.)} \pm 10.6 \text{ (syst.)} \pm 7.8\text{ (lumi.) pb},&\text{CMS}   
\end{cases}
\end{equation}
Since the uncertainty of theoretical predictions based on
next-to-leading order~(NLO) calculations~\cite{Nason:1987xz} in QCD
and next-to-leading-logarithmic~(NLL) higher-order
effects~\cite{Kidonakis:1997gm} is larger than $10\%$, higher-order
corrections have to be included in order to match the experimental
precision.  In the absence of the complete NNLO corrections that are
currently being computed,\footnote{For a status report
  see.~\cite{Bonciani:2010ue}  The result for the
  quark-antiquark initial state was obtained very
  recently.~\cite{Baernreuther:2012ws}} the theoretical precision can
be improved by including higher-order QCD corrections that are
enhanced in the partonic threshold limit, $\beta=\sqrt{1-4m_t^2/\hat
  s}\to 0$, where $\hat s$ is the partonic centre-of-mass energy.
These contributions take the form of logarithmic corrections
proportional to $(\alpha_s\log^{2,1}\beta)^n$ due to the emission of
soft gluons, and of Coulomb corrections $(\alpha_s/\beta)^n$ due to
the virtual exchange of gluons between the slowly moving top
quarks. Both corrections can be resummed to all orders in perturbation
theory, leading to a representation of the partonic cross sections for
the subprocesses $pp'\to t\bar t X$ (with $p,p'\in \{g,q,\bar q\})$ of
the form
\begin{eqnarray}
\label{eq:syst}
\hat{\sigma}_{p p'} &=& \,\hat \sigma^{(0)}_{p p'}\, 
\sum_{k=0} \,\left(\frac{\alpha_s}{\beta}\right)^{\!k} \,
\exp\Big[\underbrace{\ln\beta\,g_0(\alpha_s\ln\beta)}_{\mbox{(LL)}}+ 
\underbrace{g_1(\alpha_s\ln\beta)}_{\mbox{(NLL)}}+
\underbrace{\alpha_s g_2(\alpha_s\ln\beta)}_{\mbox{(NNLL)}}+\ldots\Big]
\nonumber\\[0.2cm]
&& \,\times
\left\{1\,\mbox{(LL,NLL)}; \alpha_s\,\mbox{(NNLL)}; 
\alpha_s^2 ,\beta^2 \,\mbox{(N$^3$LL)};
\ldots\right\}\, . 
\end{eqnarray}
Several recent  developments have made it possible to perform resummation at NNLL accuracy:
 the function $g_2$ has been computed~\cite{Beneke:2009rj} using 
 the infrared structure of massive QCD amplitudes,~\cite{Becher:2009kw} while
 the $\mathcal{O}(\alpha)$ coefficient functions~\cite{Czakon:2008cx}  and the NNLO Coulomb effects~\cite{Beneke:2009ye} became available as well.
The combined resummation of soft and Coulomb corrections has been established in.~\cite{Beneke:2010da}

\section{Results from NNLL resummation}
In~\cite{Beneke:2011mq} we have performed the combined NNLL
resummation of soft and Coulomb effects using the momentum-space
approach to soft-gluon resummation~\cite{Becher:2006nr} and results
for the higher-order Coulomb corrections,~\cite{Beneke:1999qg}
including would-be bound-state contributions to the cross section.
We extend these results in table~\ref{tab:res} by providing predictions for the LHC at a centre-of-mass energy of $8$~TeV in addition to the results for Tevatron and the LHC at $7$ and $14$ TeV. 
Results for different values of $m_t$ at $8$~TeV (updated to include the NNLO $q\bar q$ partonic cross section~\cite{Baernreuther:2012ws}), as well as for hypothetical heavy quarks will be presented elsewhere.~\cite{Beneke:2012}
For comparison, the table also includes the NLO cross section~\cite{Nason:1987xz} and the approximate NNLO results~\cite{Beneke:2009ye} obtained by expanding the resummed corrections to $\mathcal{O}(\alpha_s^2)$. 
The theoretical uncertainty of the approximate NNLO results includes an estimate of the unknown constant NNLO contribution to the cross section; the NNLL uncertainties include in addition an estimate of higher-order ambiguities based on comparing different NNLL implementations and expansions to N${}^3$LO accuracy as discussed in detail in.~\cite{Beneke:2011mq}
Compared to the NLO results, the NNLL corrections increase the cross section by $8\%$ at the Tevatron and $1-3\%$ at the LHC.
The main  effect of the NNLL corrections is included in the NNLO${}_{\text{app}}$ result, with further higher-order corrections of about $2\%$  at the Tevatron, and $\lesssim 1\%$ at the LHC.
The  NNLO${}_{\text{app}}$  and NNLL results include two-loop Coulomb and soft/Coulomb interference effects  of the order of $1-2\%$, while Coulomb corrections beyond NNLO and bound-state contributions of the order of $0.5\%$~\cite{Beneke:2011mq} are  included in addition at NNLL.
\begin{table}[t!]
\begin{center}
\begin{tabular}{|l|c|c|c|c|}
\hline
$\sigma_{t \bar{t}}$[pb]&  Tevatron
& LHC ($\sqrt{s}=$7 TeV) & LHC ($\sqrt{s}=$8 TeV) &  LHC ($\sqrt{s}=$14 TeV) \\
\hline
\hline
NLO & $6.68^{+0.36+0.51}_{-0.75-0.45}$ & $158.1^{+18.5+13.9}_{-21.2-13.1}$
& $226.2^{\,+27.8+19.1}_{\,-29.7-17.8}$
 & $884^{+107+65}_{-106-58}$ \\
\hline
NNLO$_\text{app}$ & $7.06^{+0.27+0.69}_{-0.34-0.53}$ & $161.1^{+12.3+15.2}_{-11.9-14.5}$& 
$230.0^{\,+16.7 +20.5}_{\,-15.7-19.8} $ &
 $891^{+76+64}_{-69-63}$ \\
\hline
NNLL & $7.22^{+0.31+0.71}_{-0.47-0.55}$ & $162.6^{+7.4+15.4}_{-7.5-14.7}$& 
$231.9^{\,+10.5 +20.8}_{\,-10.3-20.1}$& $896^{+40+65}_{-37-64}$ \\
\hline
\end{tabular}
\end{center}
\caption{$t \bar{t}$ cross section at Tevatron and LHC in various approximations, for $m_t=173.3\,$GeV using the MSTW08 PDFs. The first error denotes 
the total theoretical uncertainty, the second the $90\%$ c.l. PDF+$\alpha_s$ uncertainty. }
\label{tab:res} 
\end{table} 

\begin{figure}[t!]
\begin{center}
\includegraphics[width=0.52 \linewidth]{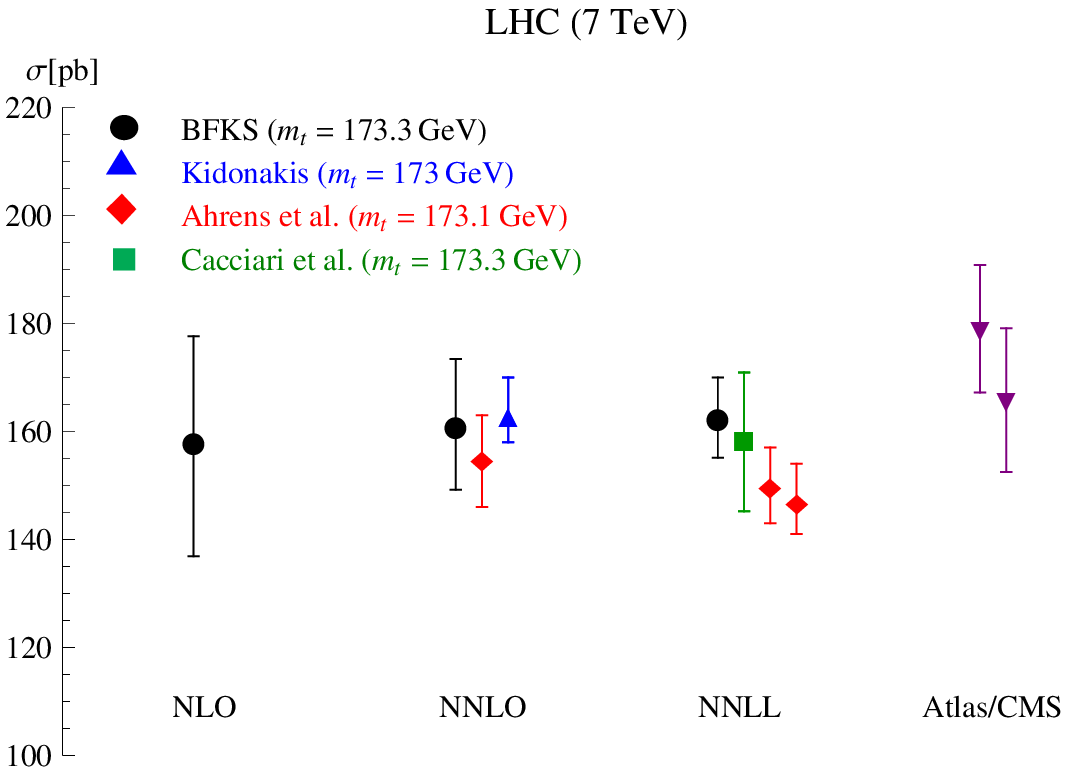}
 \includegraphics[width=.42\textwidth]{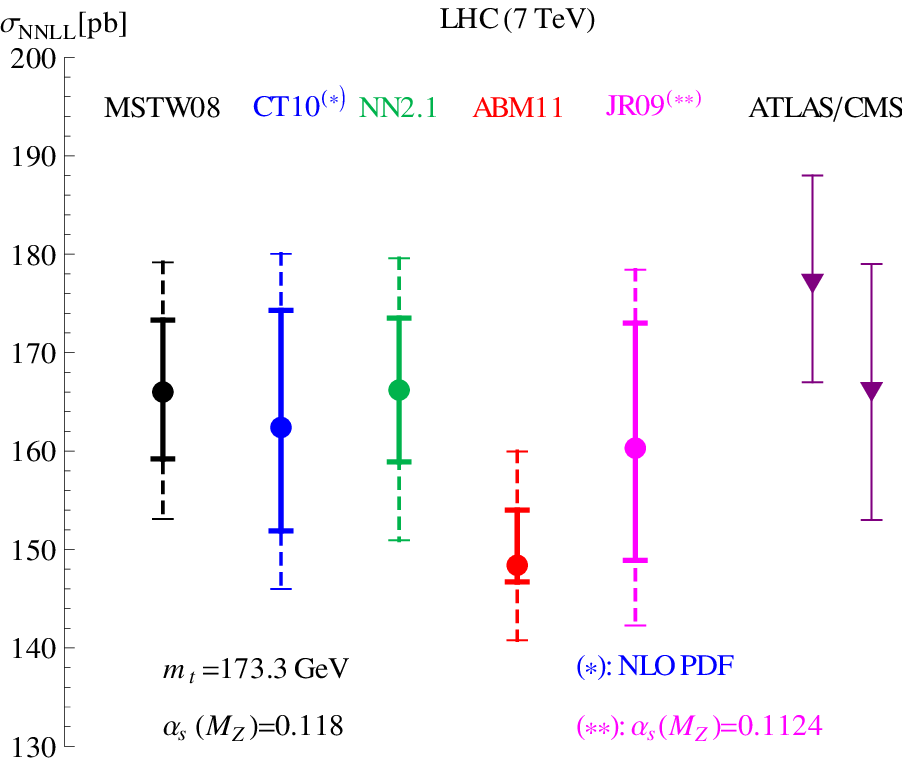}  
\end{center}
\caption{Left: Comparison of different NNLO and NNLL predictions, see the text for explanation and references.
The error bands include theoretical uncertainties, but no PDF+$\alpha_s$ errors.
Right: NNLL predictions for different PDF sets with fixed  $\alpha_s(M_Z)=0.118$. The inner (solid) error bar denotes the $68\%$ confidence level PDF$+\alpha_s$ error, the outer (dashed) error bar includes in addition the uncertainty of the NNLL cross-section calculation.
Both figures include also the most recent experimental measurements that assume  $m_t=172.5$~GeV.}
\label{fig:comp}
\end{figure}

In the left panel of figure~\ref{fig:comp} we compare our results (denoted by
black circles) for the LHC at $\sqrt s=7$~TeV to predictions by other groups
and experimental measurements.  The NNLL resummation of soft-gluon corrections
using the traditional Mellin-space approach~\cite{Cacciari:2011hy} (denoted by a green square) differs from our
results in the treatment of constant NNLO and power-suppressed terms, in
addition to the different resummation formalism.  Further results have been
obtained by integrating NNLL or approximate NNLO predictions for
invariant-mass or $p_T$-distributions.~\cite{Kidonakis:2010dk,Ahrens:2011px}
These calculations (denoted by a blue triangle/red diamonds) include some
power suppressed contributions in $\beta$, but not the NNLO potential terms.~\cite{Beneke:2009ye}  All the
approximations~\cite{Cacciari:2011hy,Kidonakis:2010dk,Ahrens:2011px} differ
from ours by neglecting the higher-order Coulomb effects.  While the
predictions agree within the quoted uncertainties at the
LHC,~\footnote{Somewhat larger discrepancies are found at the Tevatron, see
  for instance.~\cite{Beneke:2011mq}} the different central values indicate
the ambiguities inherent in threshold approximations and illustrate the
possible impact of a full NNLO calculation.

In addition to the top-quark mass and the strong coupling, the top-pair cross section depends also on the parton distribution functions~(PDFs). Due to the dominant gluon fusion channel, uncertainties in the determination of the gluon PDF  have a large impact on the cross section at the LHC and constitute the main theoretical uncertainty.
This is illustrated in  the right panel of figure~\ref{fig:comp} where the results of the MSTW08NNLO PDF~\cite{Martin:2009iq} used as our default are compared to the NNLO PDF sets by NNPDF2.1,~\cite{Ball:2011uy}  ABM11~\cite{Alekhin:2012ig} and JR09VR.~\cite{JimenezDelgado:2008hf} As illustration, results from the NLO PDF CT10~\cite{Lai:2010vv} are also included. In this comparison,  a common central value $\alpha_s(M_Z)=0.118$ of the strong coupling~\footnote{With the exception of the JR09 PDF where $\alpha_s(M_Z)=0.1124$ is used.} is employed instead of the
 best-fit values used as default by the PDF sets (note that the  MSTW08NNLO default  $\alpha_s(M_Z)=0.1171$  is used in the left panel of Figure~\ref{fig:comp}). 
For the common $\alpha_s$-value, the predictions of most PDFs agree within the $68\%$ confidence level of the PDF$+\alpha_s$ uncertainty (denoted by the inner, solid error bar), but it is also seen that the spread is larger than estimated by the uncertainty of a single PDF set.

As an application of a precise theoretical prediction of the total
top-pair cross section, the top-quark mass can be extracted in a
theoretically well defined mass definition from the measured cross
section, assuming the latter is free of new-physics
contributions.~\cite{Langenfeld:2009wd,Abazov:2011cq} Using our NNLL predictions
discussed above, we have estimated that the top-quark pole mass could
be extracted with an accuracy of $\pm 5$~GeV from the currently
available ATLAS data on the total cross section, and our
result~\cite{Beneke:2011mq} $m_t = 169.8^{+4.9}_{-4.7}$ GeV is
compatible with the direct mass determination $m_t=173.2\pm 0.8$~GeV
at the Tevatron.~\cite{Lancaster:2011wr} A CMS
analysis~\cite{Aldaya:2012tv} of a cross section measurement with a
smaller central value but larger uncertainty, $\sigma_{t\bar t}=169.90
\pm 3.9$~(stat.)~$\pm 16.3$~(syst.)$\pm 7.6$~(lumi.) pb,
obtained the
comparable result  $m_t = 170.3^{+7.3}_{-6.7}$~GeV  using the calculation of.~\cite{Langenfeld:2009wd}

\section{Summary and outlook}
We have reviewed the results of the
NNLL resummation of soft and Coulomb-gluon corrections performed in~\cite{Beneke:2011mq} 
and extended them by including predictions  for the LHC at $\sqrt s=8$~TeV and for different PDF sets. Our results are
 in good agreement with experimental measurements.
We plan to make our calculation available in form of a public program in the near future.~\cite{Beneke:2012}
 We have 
discussed the impact of the uncertainty of current PDF sets on the cross-section predictions. The prospects to constrain the gluon PDF by the top-pair cross-section measurement will be investigated in.~\cite{Beneke:2012}
 It has been  estimated that the top-quark pole mass could be measured with a precision of $\pm 5$~GeV from current LHC cross-section measurements.

\section*{References}

\providecommand{\href}[2]{#2}\begingroup\raggedright

\end{document}